\documentclass{appolb}
\usepackage{epsfig}
\usepackage{amsmath,amssymb,bm}

\newcommand{\be}{\begin{equation}}
\newcommand{\ee}{\end{equation}}
\newcommand{\bea}{\begin{eqnarray}}
\newcommand{\eea}{\end{eqnarray}}

\newcommand{\vecv}{\bm v}

\newcommand{\vecnu}{\bm \nu}

\begin{document}
\title{Astrophysics of dense quark matter in compact stars%
\thanks{Presented at the XXVI Max Born Symposium {\it ``Three days of 
        strong interactions''}, Wroc\l aw, Poland,
9-11 July, 2009.}%
}
\author{Armen Sedrakian
\address{
Institute for Theoretical Physics,
J. W. Goethe-University,\\D-60438 Frankfurt am Main, Germany
}
}

\maketitle
\begin{abstract}
Massive neutron stars may harbor deconfined quark 
matter in their cores. I review some recent work on the 
microphysics and  the phenomenology of compact stars 
with cores made of quark matter. This includes the 
equilibrium and stability of non-rotating and rapidly 
rotating stars, gravitational radiation from 
deformations in their quark cores, neutrino radiation and 
dichotomy of fast and slow cooling, and pulsar radio-timing 
anomalies.
\end{abstract}
  
\section{Introduction}
\label{sec:Intro}

Because of the quark substructure of nucleons predicted by quantum 
chromodynamics (QCD), nuclear matter will undergo a phase transition 
to quark matter if squeezed to sufficiently high densities. In the 
quark matter phase the ``liberated'' quarks occupy continuum states 
which, in the low-temperature and high-density regime, arrange 
themselves in a Fermi sphere. The Fermi-sphere determines the form 
of low-lying excitation spectrum of quark matter, which resembles that 
of less exotic low-temperature systems found in condensed matter 
(\eg electron gas or ultracold atomic vapor) or hadronic physics 
(\eg nuclear or neutron matter). In analogy to these system the 
attractive interaction between quarks, mediated by the gluon exchange 
(which is responsible for the bound state spectrum of QCD, \eg, the 
nucleon and mesons)  leads to quark superconductivity and 
superfluidity (color superconductivity) via the Bardeen-Cooper-Schrieffer 
mechanism~\cite{Bailin:1983bm}.
Furthermore, under stellar conditions the pairing between the two 
light flavors of quarks occurs at finite isospin chemical potential,
\ie, when the Fermi surfaces of up and down quarks are shifted apart 
by an amount which is of the same order of magnitude as the gap
in the quasiparticle spectrum. Under these conditions the pairing 
between the fermions persists, but the actual pairing pattern  may 
be significantly different from that of the BCS and is likely to 
involve breaking of the spatial symmetries by the condensate order 
parameter.

During the last decade there has been a substantial progress in 
understanding of pairing in two-component asymmetric superconductors. 
Firstly, new developments on (variations of) the 
Larkin-Ovchinnikov-Fulde-Ferrell 
(LOFF) phase revealed novel lattice structures of the order 
parameter~\cite{Bowers:2002xr}.
Secondly, it has been suggested that such systems may actually 
phase separate into normal and superconducting domains~\cite{Caldas} 
or the pairing may require changes in the shapes of the Fermi 
surfaces~\cite{Muther:2002mc}. Controlled experiments on two-component
cold atomic vapors with mismatched Fermi surfaces show that the phase 
separation scenario is at work. Whether this is the 
case in the related systems such as the isospin asymmetric nuclear 
matter or deconfined quark matter is still unclear. Therefore,
a major goal of astrophysics of dense matter is to
find and to quantify the manifestations of dense phases in the observable 
properties of compact stars. Here we review some recent progress
towards this goal.

\section{Color superconducting phases with 
                                broken spatial symmetries}
\label{sec:sup_phases}

The quark-quark interaction is strongest for pairing between up and 
down quarks which are antisymmetric in color~\cite{Bailin:1983bm}.
At intermediate densities 
quark matter is composed of up and down quarks, while strange quarks
can appear in substantial amounts at higher densities. 
The leading candidate phase at these densities is the so-called 2SC 
phase, which is characterized by the order parameter 
$
\Delta \propto \langle
\psi^T(x)C\gamma_5\tau_2\lambda_2\psi(x)\rangle ,
$ where
$\tau_2$ is the Pauli matrix in the isospin state, $\lambda_2$
is the Gell-Mann matrix in the color space, $C = i\gamma^2\gamma^0$ 
is the matrix of charge conjugation. 
The isospin chemical potential is  
determined by the $\beta$-equilibrium condition 
$\mu_d-\mu_u =\mu_e\sim 100$ MeV among $d$ and $u$ quarks and electrons, 
where $\mu_i$ with $i=d,\, u,\, e$ are the chemical potentials. The
Fermi spheres of up and down quarks are shifted apart by this amount.  
Thus, the cross-flavor pairing must overcome the disruptive 
effect of mismatched Fermi surfaces.
Furthermore, if the physical strange quark mass is close to its 
current mass $\sim 100$ MeV, electrons may be gradually replaced 
by strange quarks, which again will disfavor the cross-flavor pairing. 

A superconducting phase with asymmetric cross-species pairing needs to 
optimize the overlap between the Fermi surfaces (which is perfect in the 
symmetric BCS state) to attain the  maximum possible condensation energy. 
There are a number of ways to achieve this: (a) condensate that 
carries non-zero momentum with respect to 
some fixed frame. The penalty in the energy budget 
for the (always positive) kinetic energy of condensate 
motion is compensated by the gain in the (negative) condensation 
energy. In three-flavor quark matter the 
face-centered cubic (fcc) lattices were identified as having particularly 
low free energy~\cite{Bowers:2002xr} in the regime 
where the Ginzburg-Landau (GL) theory is applicable. 
The spatial form of the condensate beyond the GL 
regime is not known, therefore current treatments beyond this 
regime require some simplifying 
assumptions~\cite{Nickel:2008ng,Sedrakian:2009kb}. (b)
A more symmetric superconducting phase exploits 
deformations of Fermi spheres as the mechanism of restoring the coherence 
needed for pairing at the cost of kinetic energy loss caused by these 
deformations.  The deformed Fermi surface phase requires minimal 
breaking of spatial symmetry and is more symmetric than the lattice 
phases above~\cite{Muther:2002mc}. 

In this review we will be concerned mainly with the 2SC and 
the crystalline color superconducting (CCS) phases. We expect that 
the phenomenology of related phases with cross-flavor pairing 
is essentially the same.

\section{Equilibrium and stability}
\label{sec:equilibrium}
\begin{figure}[tb]
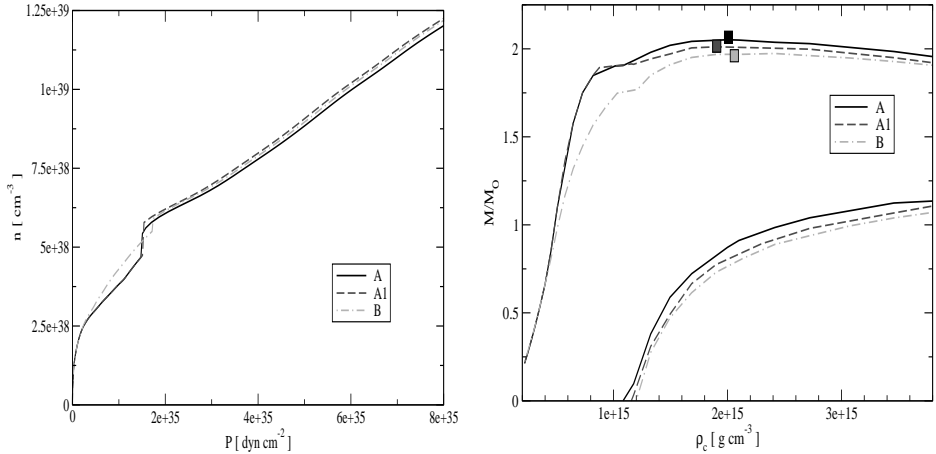

\hskip 0.1 cm
\begin{center}
\epsfig{figure=n_P,height=6.cm,width=6.cm,angle=0}
\hskip 0.2cm
\epsfig{figure=M_density,height=6.cm,width=6.cm,angle=0}
\vskip 0.13cm
\caption{
{\it Left panel}.
Number density versus pressure for three models.
For the models A and A1 the  nuclear (low density) equation of
state is the same; for the models A and B the  quark (high density)
equation of state is the same. {\it Right panel}.
Dependence of the total stellar mass and the mass of the quark core
in units of solar mass $M_{\odot}$
on the central density for non-rotating configurations. 
The lower set of curves represents the masses of the 
CCS quark cores, the upper set - the total masses of the
configurations. The maximal masses are marked with boxes.
}
\label{fig:M_density}
\end{center}

\end{figure}

The central question, which we address in this section, 
is whether the equation of state of matter at
high densities admits stable configurations of self-gravitating
objects in General Relativity featuring deconfined quark matter.
The deconfinement phase transition from baryonic to quark matter 
leads to a softening of the equation of state which could lead to 
an instability towards a collapse into a black hole. 
In ref.~\cite{Ippolito:2007hn}
the quark phase was studied in the Nambu--Jona-Lasinio (NJL) model, 
which is a low-energy non-perturbative
approximation to  QCD, that is anchored in the low-energy
phenomenology of the hadronic spectrum. While  dynamical symmetry
breaking, by which quarks acquire mass, is incorporated in this model,
it lacks confinement.
Physically, the true nuclear equation state must go over to some 
sort of quark equation of state at some density if the deconfinement
has to take place in nature.
If pressure vs chemical potential curves for chosen equations of state 
of nuclear and quark matter do not cross,
then the models are incompatible
in the sense that they can not describe the desired transition between 
the nuclear and quark matter. The low-density equation of state of nuclear 
matter and the high-density  equation of state of 
CCS matter were matched in ref.~\cite{Ippolito:2007hn}
at an interface via the Maxwell construction.
The phase with largest pressure is the one that is realized at a given
chemical potential. Thus, at 
the deconfinement phase transition there is a jump in the density
at constant pressure as illustrated in Fig.~\ref{fig:M_density}, left 
panel. The low-density nuclear equations of state are based 
on the Dirac-Bruckner-Hartree-Fock approach
and are the most hard equations of state in our 
collection~\cite{WEBER_BOOK,Sedrakian:2006mq}.
It should be noted that the matching with softer
equations of state can be enforced by varying the normalization 
of the pressure (``effective bag constant'') of the quark matter, 
in which case the density of deconfinement is a free parameter.

The spherically symmetric solutions of Einstein's
equations for self-gravitating fluids are given by the well-known
Tolman-Oppenheimer-Volkoff equations~\cite{WEBER_BOOK}. A generic feature
of these solutions is the existence of a maximum mass for any
equation of state; as the central density is increased beyond the value 
corresponding to the maximum mass, the stars
become unstable towards collapse to a black hole. A criterion
for the stability of a sequence of configurations is the requirement
that the derivative $dM/d\rho_c$ should be positive (the mass should
be an increasing function of the central density). 

For configurations constructed from a purely nuclear equation 
of state the stable sequence extends up to a maximum mass 
of the order 2 $M_{\odot}$ (Fig. \ref{fig:M_density}), 
this large value being a consequence of hardness 
of the equation of state.
The hybrid configurations branch off from the nuclear
configurations when the central density reaches that of the
deconfinement phase transition. It is seen that a stable branch 
of hybrid stars emerges in the range of central densities 
$1.3 \le \rho_c\le 2.5 \times 10^{15}$ g cm$^{-3}$. 
The masses of the CCS quark cores cover the range  $0\le M_{\rm core}
/M_{\odot}\le 0.75-0.88$ for central densities $1.3 \times 
10^{15}\le \rho_c\le 2\times 10^{15}$. Thus, 
the quark core mass ranges from one third to about the half of the 
total stellar mass.

Millisecond neutron stars can rotate at frequencies which 
are close to the limiting orbital Keplerian frequency at 
which  mass shedding from the equatorial plane starts. 
The Keplerian frequency sets an upper limit on the rotation 
frequency, since other (less certain) mechanisms, such
as secular instabilities, could impose lower limits on
the rotation frequency. The mass versus central density dependence
of compact stars rotating at the Keplerian frequency is 
similar to that for non-rotating stars  with the scales 
for mass shifted to larger values~\cite{Ippolito:2007hn}.
The increase in the maximum mass for stable
hybrid configurations (in solar mass units) is
$2.052\to 2.462$ for the model A,  $2.017\to 2.428$ and $2.4174$ for
the model A1 (there are two maxima) and  $1.981\to 2.35$ 
for the model B~\cite{Ippolito:2007hn}.

Thus, a new branch of {\it stable hybrid configurations} 
with CCS matter cores emerges within a broad range of central 
densities. The quark equation of state depends only on the
NJL model parameters (\ie does not contain an additional 
``bag constant''). A quite general conclusion of our analysis 
is that the nuclear equation of state needs to be {\it hard} 
to enable thermodynamical equilibrium with the
NJL model quark matter.

\section{Gravitational radiation}
\label{sec:grav_rad}
\begin{figure}[tb]
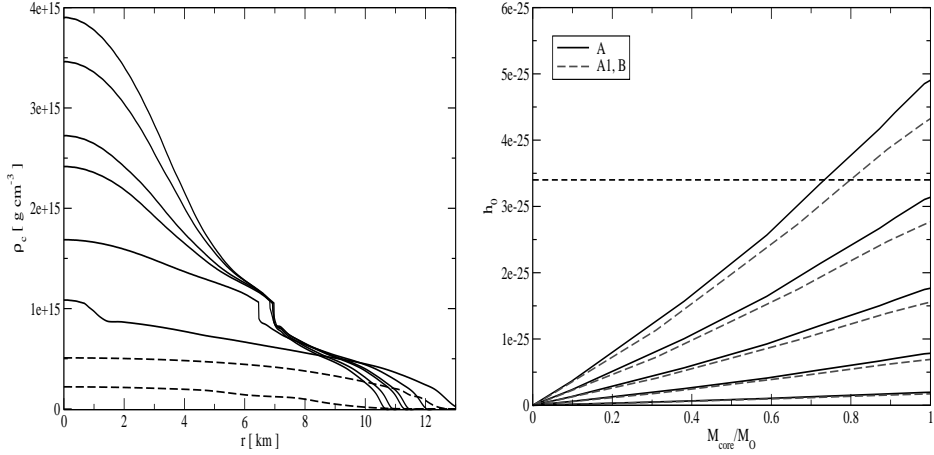

\hskip 0.1 cm
\begin{center}
\epsfig{figure=density_radius_innerhalb_spline,height=6.cm,width=6.cm,angle=0}
\hskip 0.2cm
\epsfig{figure=h_M,height=6.cm,width=6.cm,angle=0}
\vskip 0.3cm
\caption{ 
{\it Left panel.}
Dependence of density on the internal radius (model A). 
The dashed lines correspond to purely nuclear, the solid 
lines - to hybrid configurations. Density jumps at the 
phase transition to quark matter.
{\it Right panel.}
The strain of gravitational wave emission of the Crab pulsar.
Each triple of  curves corresponds to the gap  value 
(from top to bottom) $\Delta =50$, 40, 30, 20, and 10 MeV and  
the breaking strain  $\bar\sigma_{\rm max}=10^{-4}$. The horizontal 
line shows the current experimental upper limit. 
}
\label{fig:h_M_DELTA=cont_version2}
\end{center}
\end{figure}

Continuous gravitational waves emitted by non-axisymmetric 
rotating compact stars are expected to be in the 
bandwidth of current gravitational wave interferometric detectors. 
Upper limits on the strain of gravitational waves 
from a selection of radio pulsars were set 
by the LIGO collaboration~\cite{LIGO_S5_CRAB}.
Gravity waves arise from time-dependent 
quadrupole deformations of masses. Therefore, rotating isolated 
neutron stars will emit gravitational radiation if their 
mass distribution is non-axisymmetric with respect to their 
rotation axis. The axial symmetry of the star's core can be 
broken by the solid deformations in CCS matter, whose 
shear modulus $\mu_{\rm shear}$
was computed in ref.~\cite{Mannarelli:2008zz}. Initially, 
the gravitational wave emission from a quark star made of 
uniform-density incompressible CCS matter was estimated 
by Lin~\cite{Lin:2007}.
Haskell et al.~\cite{Lin:2007} estimated core deformations and 
associated with them constraints on the QCD parameters for sequences of 
1.4 $M_{\odot}$ and $R=10$ km stars composed of CCS incompressible 
quark cores and hadronic shells obeying $n=1$ polytropic equation of 
state. 
Gravitational radiation of models based on realistic equations of 
state were considered in ref.~\cite{Knippel:2009st}. 
One key difference to the previous models is that realistic hybrid  
configurations have masses that are close to the maximum sustainable 
mass $M_{\rm max}\simeq 2M_{\odot}$. Furthermore, the quadrupole moment 
of the quark core was computed from realistic density 
profile, which differs substantially from constant density 
profile of an incompressible fluid, see Fig.~\ref{fig:h_M_DELTA=cont_version2}.
The characteristic strain amplitude of gravitational waves 
emitted by a triaxial star rotating about its principal axis is  
$
h_0=16\pi^2 G c^{-4} \epsilon I_{zz}\nu^2r^{-1},
$
where $\nu$ is the star's rotation frequency, $r$ is the distance to the 
observer, $\epsilon=(I_{xx}-I_{yy})/I_{zz}$ is the equatorial ellipticity, 
$I_{ij}$ is the tensor of moment of inertia, $G$ is the
gravitational constant, $c$ is the speed of light. The elastic deformations 
are assumed to be small perturbation on the background equilibrium of the 
star.
Fig.~\ref{fig:h_M_DELTA=cont_version2} displays the strain of 
gravitational wave emission as a function of the core mass 
$M_{\rm core}$  and the current upper limit for the 
Crab pulsar $h_0 < 3.4 \times 10^{-25}$, which is rotating 
at the frequency $\nu = 29.6$ Hz  at the distance 
2 kpc~\cite{LIGO_S5_CRAB}. The value of the
breaking strain of CCS core is $\bar\sigma_{\rm max}
=10^{-4}$. The dependence of the stain amplitude on the microscopic 
parameters follows from the chain $h_0\propto Q_{\rm max}\propto 
\mu_{\rm shear}\propto \bar\sigma_{\rm max}\Delta^2$, whereby 
$Q_{\rm max}$ is the maximal quadrupole moment of the CCS core. 
The strain of gravitational wave emission is close to the upper 
limit for $\Delta = 50$ MeV.  If, however, the gaps are small 
more ``optimistic'' values of $10^{-3}\le \bar\sigma_{\rm max}
\le 10^{-2}$ will be needed to generate sizeable strain amplitude.
The  Crab puslar's current limit implies $\bar\sigma_{\rm max}\Delta^2  
\sim 0.25$ MeV$^2$, assuming maximally strained matter. 
The evolutionary avenues that may lead to such maximal deformations 
are not known. Pulsars may simply preserve their initial deformations
as they cool down and solidify in the CCS state.
 We conclude that {\it massive compact stars with CCS quark cores 
are strong candidate sources of gravitational wave emission}. 
Improved upper limits on the strain amplitude, can narrow down 
the admissible range of parameters of the CCS matter in the future.

\section{Neutrino emission and cooling}
\label{sec:cooling}
Neutrino emissivities control the cooling rate of a neutron star 
during the first $10^4-10^5$ yr of their evolution. At later times 
the photon emission from the surface and the heating in 
the interior are the main factors.  Depending on the dominant 
neutrino emission process in the neutrino emission era $t\le 10^5$ 
yr the cooling could be slow (standard) or  fast 
(nonstandard).

The slow cooling scenario is based on neutrino cooling via the
modified Urca and bremsstrahlung processes. The fast cooling 
requires unconventional processes such as decays in 
pion/kaon condensates, 
direct Urca process on nucleons, hyperons, or quarks.
Phase-space arguments show that fast cooling neutrino 
processes have typically 
temperature dependences $\sim T^6$, whereas those leading to slow cooling 
require  additional phase space for the spectator particle 
and their emissivities scale as $T^8$.

An inspection of the observational data on neutron star 
surface temperatures, which is commonly presented on a 
plot of photon-luminosity (or surface temperature) vs age,
shows that the data cannot be described by a single
cooling track; a regulator is needed that will cool some 
stars faster than the others~\cite{WEBER_BOOK,Sedrakian:2006mq}.
It is reasonable to assume 
that the heavier stars cool via some fast mechanism, while 
the lighter stars cool slowly via the modified processes. 
Indeed, the  fast cooling agents operate above a certain 
density threshold. Therefore the dichotomy of observed high 
and low surface temperature could be an evidence of a fast 
process operating in massive stars.
Thus, we are led to examine the potential fast cooling processes 
in the quark cores of compact stars. The key process 
is the Urca processes (quark $\beta$-decay)
$d\to u+e+\bar\nu_e$ and $u+e\to d+\nu_e$, which is 
permitted in {\it interacting} quark matter for any asymmetry between 
$u$ and $d$ quarks.  The effect of asymmetric pairing is well 
demonstrated on the example of 2SC 
phase~\cite{Alford:2004zr,Jaikumar:2005hy}.
The result for the emissivity depends on whether the parameter
$z = \Delta/\delta\mu$ is larger or smaller than unity, 
where $\Delta$ is the gap in the 2SC phase and $\delta\mu = \mu_d-\mu_u$ 
is the shift in the chemical potentials of the up and down quarks.
For $z > 1$ all the particle modes are ``gapped'',
therefore, as the temperature is lowered, the emissivity is suppressed
(for  asymptotically low temperatures exponentially). When $z< 1 $
there are gapless modes in the quasiparticle spectrum; this implies
that the neutrino production is not affected by color superconductivity.
As a result, the superconducting quark matter cools           
at a rate comparable to the unpaired matter.  More generally, we may 
conclude that {\it the existence of ungaped segments on the Fermi 
surfaces of quarks will lead to an enhanced cooling of these stars.}
A density dependent $z$ parameter may resolve the dichotomy of 
the fast and slow cooling by providing a smooth transition between 
these extremes.

\section{Radio-timing and rotational anomalies}
\label{sec:rotation}

Pulsars are nearly prefect clocks, whose  pulsed emissions, with
a typical periodicity of seconds or less, are locked to the rotation 
period of the star. Their periods increase gradually over time, 
due to a secular loss of rotational energy. Some 
pulsars show deviations from this regularity.  
The pulsar timing anomalies divide roughly into
three types. (i)~{\it Glitches}.  These are
distinguished by abrupt increases in the rotation and spin-down 
rates of pulsars by amounts $\Delta\Omega/\Omega \sim 10^{-6}-10^{-8}$ and 
$\Delta\dot \Omega/\Omega \sim 10^{-3}$.  After a glitch, 
$\Delta\Omega/\Omega$ and $\Delta\dot \Omega/\Omega$ slowly
relax toward their pre-glitch values, on a time scale of order
weeks to years, in some cases with permanent hysteresis 
effects.
Such behavior is attributed to a component within the star 
that is only weakly coupled to the rigidly rotating normal component
responsible for the emission of pulsed radiation.  
(ii)~{\it Timing Noise}. 
These represent irregular,  stochastic deviations in 
the spin and spin-down rates that are superimposed on the 
near-perfect periodic rotation of the star.  Whether the 
superfluids are involved in the generation of timing noise 
remains unclear.
(iii)~{\it Long-Term Periodic Variabilities}.  
Observed in the timing of few pulsars, most notably PSR 
B1828-11, these deviations strongly constrain theories of 
superfluid friction, if their periodicities are 
interpreted in terms of precession. (Note however that the 
Tkachenko modes of the vortex lattice are a viable alternative 
to the precession interpretation~\cite{Noronha:2007qf}.)
The importance of the inferred
precession mode stems from the fact that it involves 
non-axisymmetric perturbations of the rotational state,
removing the degeneracy with respect to $\zeta$ ($\zeta'\simeq 0$)
that is inherent in the interpretation of post-glitch dynamics.

The macroscopic physics of neutron-star rotation and its anomalies
observed in the timing of pulsars can be described within the 
hydrodynamic theory of superfluids. At the local hydrodynamical scale, 
the rate of angular momentum transfer between the superfluid 
and normal components is determined by the equation of motion 
of a vortex line in a neutral superfluid
\be\label{FORCE_BALANCE} 
\omega_S (\vecv_S-\vecv_L)\times\vecnu + 
\zeta (\vecv_L-\vecv_N) +\zeta' 
(\vecv_L-\vecv_N)\times \vecnu  =0\,,
\ee
where $\vecv_S$ and $\vecv_N$ are the superfluid and normal 
fluid velocities, $\vecv_L$ is the velocity of the vortex,
$\vecnu$ is a unit vector along the vortex line, $\omega_S$ 
is the unit of circulation, and $\zeta$, $\zeta'$ are 
(dimensionless) friction coefficients, also known as the 
drag-to-lift ratios.  These coefficients encode the 
essential information on the microscopic processes of 
interaction of vortices with the ambient unpaired fluid. 

The potential role of quark matter in the rotational dynamics  
of compact stars is not understood yet. Some of the candidate 
phases, such as the  color-flavor-locked (CFL) phase, rotate by             
creating vortices~\cite{Iida}, which interact 
with the phonon gas within this 
phase~\cite{Mannarelli:2008je}. 
However, phonons do not interact electromagnetically with the rest 
of the star, therefore it is unclear on which timescales the CFL 
quark core will couple to the observable crust. Such a coupling 
is manifest for the 2SC and related phases, where the magnetic field 
(partially) penetrates into the superfluid region. 
While the vortices in these phases (if any) are not              
rotational in nature, their electromagnetic coupling to the vortices 
in the hadronic core and to electrons, which are shared with less 
denser phases
of the star, can impact the rotational dynamics of pulsars.

I would like to thank J. W. Clark, N. Ippolito, P. Jaikumar, B. Knippel,
D. H. Rischke, C. D. Roberts, and F. Weber for collaboration on the 
topics discussed in this review.  This work was in part supported 
by the Deutsche Forschungsgemeinschaft (Grant SE 1836/1-1)


\begin{thebibliography}{99}

\bibitem{Bailin:1983bm}
  D.~Bailin and A.~Love,
  Phys.\ Rept.\  {\bf 107}, 325 (1984);
  M.~Alford and K.~Rajagopal,
  in {\it Pairing in Fermionic Systems}, edited by A. Sedrakian, J. W. 
  Clark and M. Alford, (World Scientific, Singapore, 2006), p. 1,
  arXiv:hep-ph/0606157;
  S.~B.~R\"uster, V. Werth, M. Buballa, I. A. Shovkovy and D. H. Rischke, 
  {\it ibid.}, p. 63,
  arXiv:nucl-th/0602018;
  T.~Schafer,
  {\it ibid.}, p. 109,
  arXiv:nucl-th/0602067;
  M.~G.~Alford,
  arXiv:0907.0200 [nucl-th].


\bibitem{Bowers:2002xr}
J.~A.~Bowers and K.~Rajagopal,
Phys.\ Rev.\ D {\bf 66}, 065002 (2002) [arXiv:hep-ph/0204079];
  K.~Rajagopal and R.~Sharma,
  Phys.\ Rev.\ D {\bf 74}, 094019 (2006)
  [arXiv:hep-ph/0605316];
  J.\ Phys.\ G {\bf 32}, S483 (2006)
  [arXiv:hep-ph/0606066];
  N.~D.~Ippolito, G.~Nardulli and M.~Ruggieri,
  JHEP {\bf 0704}, 036 (2007)
  [arXiv:hep-ph/0701113].

\bibitem{Caldas}
P. F. Bedaque, H. Caldas and  G. Rupak
Phys.\ Rev.\ Lett.\  {\bf 91}, 247002 (2003);
H. Caldas, Phys.\ Rev.\ Rev.\ A {\bf 69}, 063602 (2004).


\bibitem{Muther:2002mc}
  H.~M\"uther and A.~Sedrakian,
  Phys.\ Rev.\ Lett.\  {\bf 88}, 252503 (2002)
  [arXiv:cond-mat/0202409];
  Phys.\ Rev.\  C {\bf 67}, 015802 (2003)
  [arXiv:nucl-th/0209061];
  Phys.\ Rev.\  D {\bf 67}, 085024 (2003)
  [arXiv:hep-ph/0212317];
  A.~Sedrakian,
  in {\it Superdense QCD Matter and Compact Stars}, edited by D. Blaschke
  and D. Sedrakian, (Springer, Dordrecht, 2006), p. 209,
  arXiv:nucl-th/0312053.

\bibitem{Nickel:2008ng}
  D.~Nickel and M.~Buballa,
  Phys.\ Rev.\  D {\bf 79}, 054009 (2009)
  [arXiv:0811.2400 [hep-ph]].

\bibitem{Sedrakian:2009kb}
  A.~Sedrakian and D.~H.~Rischke,
  Phys.\ Rev.\  D {\bf 80}, 074022 (2009)
  arXiv:0907.1260 [nucl-th].


\bibitem{Ippolito:2007hn}
  N.~Ippolito, M.~Ruggieri, D.~H.~Rischke, A.~Sedrakian and F.~Weber,
  Phys.\ Rev.\  D {\bf 77}, 023004 (2008) [arXiv:0710.3874 [astro-ph]].

\bibitem{WEBER_BOOK} F.~Weber,
{\it Pulsars as astrophysical laboratories for nuclear and particle 
physics},  Bristol, U.K.: Institute of Physics, 1999;

\bibitem{Sedrakian:2006mq}
  A.~Sedrakian,
  Prog.\ Part.\ Nucl.\ Phys.\  {\bf 58}, 168 (2007)
  [arXiv:nucl-th/0601086].

\bibitem{LIGO_S5_CRAB}
B.~Abbott, {\it et al.} (LIGO Scientific Collaboration)
Astrophys. J. Letters {\bf 683}, 45 (2008).



\bibitem{Mannarelli:2008zz}
  M.~Mannarelli, K.~Rajagopal and R.~Sharma,
  Prog.\ Theor.\ Phys.\ Suppl.\  {\bf 174}, 39 (2008).

\bibitem{Lin:2007}
L.-M.~Lin,  Phys.\ Rev.\  D {\bf 76}, 081502(R) (2007);
B.~Haskell, N.~Andersson, D.~I.~Jones, and L.~Samuelsson,
Phys.\ Rev.\  Lett. {\bf 99}, 231101 (2007).


\bibitem{Knippel:2009st}
  B.~Knippel and A.~Sedrakian,
  Phys.\ Rev.\  D {\bf 79}, 083007 (2009)
  [arXiv:0901.4637 [astro-ph.SR]].

\bibitem{Alford:2004zr}
  M.~Alford, P.~Jotwani, C.~Kouvaris, J.~Kundu and K.~Rajagopal,
  Phys.\ Rev.\ D {\bf 71}, 114011 (2005)
  [arXiv:astro-ph/0411560];
  R.~Anglani, G.~Nardulli, M.~Ruggieri and M.~Mannarelli,
  Phys.\ Rev.\ D {\bf 74}, 074005 (2006)
  [arXiv:hep-ph/0607341].

\bibitem{Jaikumar:2005hy}
  P.~Jaikumar, C.~D.~Roberts and A.~Sedrakian,
  Phys.\ Rev.\  C {\bf 73}, 042801(R) (2006)
  [arXiv:nucl-th/0509093].





\bibitem{Sedrakian:1998vi}
  A.~Sedrakian, I.~Wasserman and J.~M.~Cordes, 
  Astrophys. J. {\bf 524}, 341 (1999),
  arXiv:astro-ph/9801188;
  B.~Link,
  Astrophys.\ Space Sci.\  {\bf 308}, 435 (2007);
  T.~Akgun, B.~Link and I.~Wasserman,
  Mon.\ Not.\ Roy.\ Astron.\ Soc.\  {\bf 365}, 653 (2006)
  [arXiv:astro-ph/0506606].
  B.~Link,
  arXiv:astro-ph/0608319.

\bibitem{Noronha:2007qf}
  J.~Noronha and A.~Sedrakian,
  Phys.\ Rev.\  D {\bf 77}, 023008 (2008)
  [arXiv:0708.2876 [astro-ph]].



\bibitem{Iida} K.~Iida and G.~Baym,
               Phys.\ Rev.\  D {\bf 66}, 014015 (2002).

\bibitem{Mannarelli:2008je}
  M.~Mannarelli, C.~Manuel and B.~A.~Sa'd,
  Phys.\ Rev.\ Lett.\  {\bf 101}, 241101 (2008)
  [arXiv:0807.3264 [hep-ph]].


\end{thebibliography}
\end{document}